\documentclass{PoS}

\title{A KAT--7 view of a low-mass sample of galaxy clusters}

\ShortTitle{A KAT--7 view a low-mass sample of galaxy clusters}

\author{\speaker{G.~Bernardi}$^{1,2}$, T.~Venturi$^3$, R.~Cassano$^3$, D.~Dallacasa$^4$, G.~Brunetti$^3$, V.~Cuciti$^{3,4}$, M.~Johnston--Hollitt$^5$, N.~Oozeer$^{1,6,7}$ \& O.M.~Smirnov$^{2,1}$\\
        $^1$SKA SA, 3rd Floor, The Park, Park Road, Pinelands, 7405, South Africa\\
	$^2$Department of Physics and Electronics, Rhodes University, PO Box 94, Grahamstown, 6140, South Africa\\
	$^3$INAF - Osservatorio di Radioastronomia, via Gobetti 101, 40129 Bologna, Italy\\
	$^4$Dipartimento di Fisica e Astronomia, Universit\'a di Bologna, viale Berti Pichat 6/2, I-40127 Bologna, Italy\\
	$^5$School of Chemical \& Physical Sciences, Victoria University of Wellington, Wellington 6140, New Zealand\\
	$^6$African Institute for Mathematical Sciences, 6-8 Melrose Road, Muizenberg 7945, South Africa\\
	$^7$Centre for Space Research, North-West University, Potchefstroom 2520, South Africa\\
	E-mail: \email{gbernardi@ska.ac.za}}

\abstract{Radio observations over the last two decades have provided evidence that diffuse synchrotron emission in the form of megaparsec-scale radio halos in galaxy clusters is likely tracing regions of the intracluster medium where relativistic particles are accelerated during cluster mergers. In this paper we present results of a survey of 14 galaxy clusters carried out with the 7-element Karoo Array Telescope at 1.86 GHz, aimed to extend the current studies of radio halo occurrence to systems with lower masses (M$_{\rm 500} > 4\times10^{14}$~{\rm M}${_\odot}$). We found upper limits at the $0.6 - 1.9 \times 10^{24}$~Watt~Hz$^{-1}$ level for $\sim 50\%$ of the sample, confirming that bright radio halos in less massive galaxy clusters are statistically rare.}

\FullConference{EXTRA-RADSUR2015 (*)\\
		20--23 October 2015\\
		Bologna, Italy

                \bigskip
                \hrule
                \bigskip

                \textnormal{(*) This conference has been organized
                  with the support of the Ministry of Foreign Affairs
                  and International Cooperation, Directorate General
                  for the Country Promotion (Bilateral Grant Agreement
                  ZA14GR02 - Mapping the Universe on the Pathway to
                  SKA)}
}

\begin{document}
%%%%%%%%%%%
\section{Introduction}
\label{intro:Sect}
%%%%%%%%%%%
Giant Radio Halos (RHs) are Mpc-scale, diffuse radio sources with low surface brightness that are present in the central region of a certain number of galaxy clusters. Great progress has happened over the last decade on both the theoretical (see Brunetti \& Jones 2014 for a recent review) and observational (see Feretti et al. 2012 for a recent review) study of the nature and formation of giant RHs, indicating that they are the product of turbulent re-acceleration of particles in the intracluster medium due to cluster mergers. This picture, recently supported by the discovery of a bimodal distribution of clusters in the P$_{\rm 1.4} - {\rm M}_{\rm 500}$\footnote{P$_{\rm 1.4}$ is the radio halo power at 1.4~GHz and M$_{500}$ is the total cluster mass  within the radius $R_{500}$, defined as the radius corresponding to a total density contrast $500\rho_c(z)$, where $\rho_c(z)$ is the critical density of the Universe at  the cluster redshift.} plane (Basu et al. 2012, Cassano et al. 2013, Cuciti et al. 2015), is, however, essentially supported by observations of massive clusters M$_{500} \ge 6 \times 10^{14} \, {\rm M}_{\odot}$ and the re-acceleration model predicts that giant RHs should become rarer in less massive systems. Observations of less massive galaxy clusters are therefore crucial to test the predictions of RH formation models.\\
In this paper we summarize the results of a pilot observational study aimed to populate the faint end of the P$_{\rm 1.4}$--${\rm M}_{\rm 500}$ correlation with the newly commissioned 7--element Karoo Array Telescope (KAT--7).

\section{Observations and results}
\label{sec:obs}

A sample of 18 nearby clusters was selected from the Planck SZ Cluster Catalogue (Planck Collaboration, 2014) with the following criteria:
\begin{itemize}
\item[] M$_{\rm 500} > 4 \times 10^{14}$ M$_{\odot}$ mass range;
\item[] $0.05 < {\rm z} < 0.11$ redshift interval; 
\item[] $\delta < 0^{\circ}$.
\end{itemize}
Radio observations were only available for four targets of the sample and we eventually carried out new observations for 14 clusters (see Table~\ref{tab:logs}) with KAT--7. KAT--7 is an array of 7 12~m--diameter antennas, distributed in a randomized configuration that maximizes the $uv$ coverage in $\sim$~4~hours. Baselines span a range between 26~m to 185~m, giving an angular resolution of $\sim 2.5$~arcminutes at the central frequency of 1.86~GHz. Data reduction was carried out in a standard fashion and we refer the reader to Bernardi et al. (2016) for the details. The achieved sensitivity ranges between 0.3 and 1~mJy~beam$^{-1}$ depending upon the target.\\
%%%%%%%%% Begin Table 1
%
\begin{table*}
\caption[]{The low--mass cluster sample. Clusters are listed in order of decreasing $M_{500}$ values. The coordinates are referred to the X--ray cluster centre. For Triangulum Australis cluster we used KAT--7 archive data at 1.33~GHz.}
\begin{center}
\begin{tabular}{lccc}
\hline
\hline\noalign{\smallskip}
Cluster name		& z		& RH power and upper limits	&  M$_{\rm 500}$		\\
			&		&				& (10$^{14}$M$_{\odot}$)	\\
\hline\noalign{\smallskip}
%\hline
Triangulum   		& 0.051		& \,\,23.73 		& 7.91 \\
A\,3266      		& 0.059 	& -- 			& 6.71 \\
RCXJ\,1407.8--5100	& 0.097 	& -- 			& 6.52 \\
A\,3628 	      	& 0.105	      	& --		      	& 6.49 \\
A\,3827 	      	& 0.098		& --		      	& 5.93 \\
RXCJ\,1358.9--4750  	& 0.074 	& $<24.00$ 		& 5.44 \\
A\,644        		& 0.070 	& $<23.95$		& 4.70 \\
A\,3921       		& 0.094 	& $<24.22$ 		& 4.34 \\
A\,3911       		& 0.097 	& -- 			& 4.31 \\
A\,550        		& 0.099 	& $<24.27$ 		& 4.23 \\
PSZ1G\,018.75+23.57 	& 0.089 	& \,\,24.15 		& 4.21 \\
A\,3158       		& 0.059 	& $<23.79$ 		& 4.20 \\
A\,3822       		& 0.076 	& $<24.02$ 		& 4.18 \\
A\,3695       		& 0.089 	& -- 			& 4.06 \\
A\,1650       		& 0.085 	& $<24.12$ 		& 4.00 \\
\hline\noalign{\smallskip}
\end{tabular}
\end{center}
\label{tab:logs}
\end{table*}
%%%%%%%%%%%%% end Table 1
%%%%%%%%%%%%%%%%%%%%%%%%%%%%%%%%%%%  Begin Figure 2
\begin{figure}
\centering
\includegraphics[width=1\columnwidth]{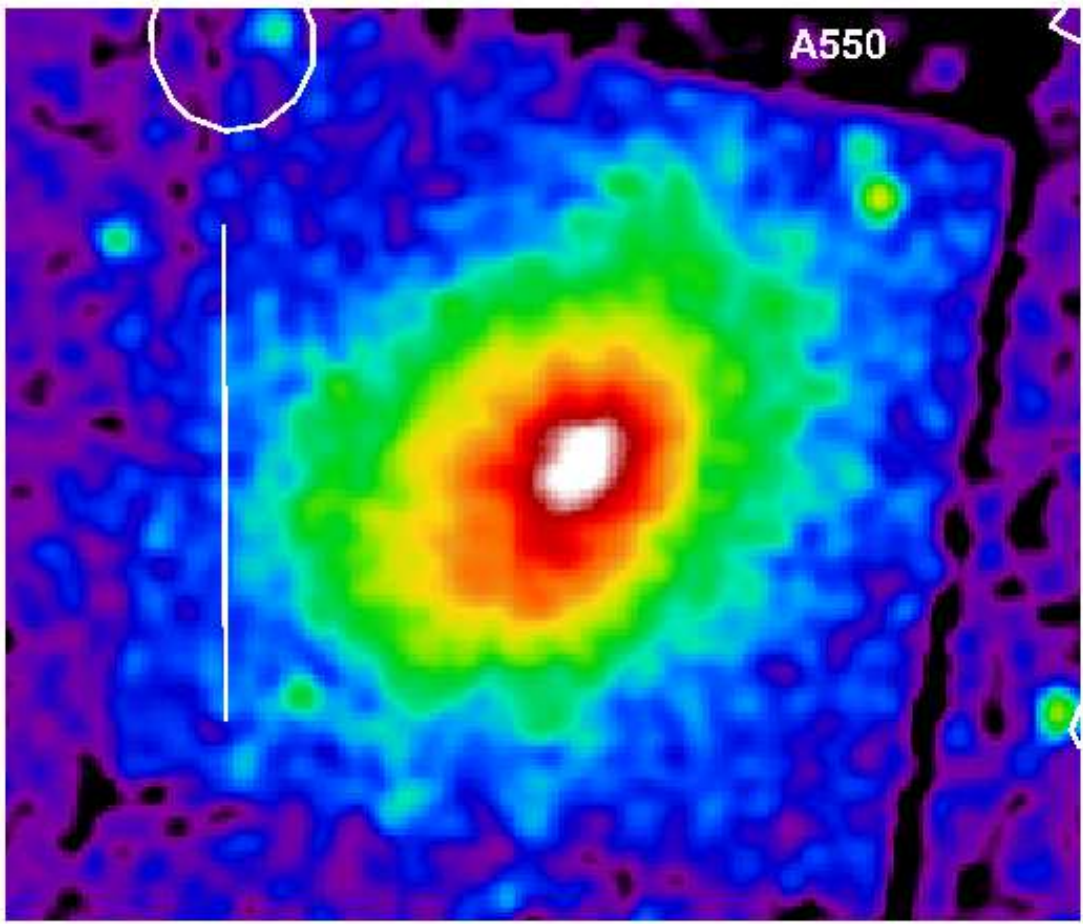}
\caption{1.86~GHz KAT--7 radio contours of the A\,550 cluster overlaid on the X--ray {\it XMM--Newton} images as an example of target that does not show any diffuse radio emission corresponding to the cluster centre (from Bernardi et al. 2016). Radio contours are drawn at -2.5, 2.5, 10 and 40~mJy~beam$^{-1}$ with positive (negative) contours drawn using solid (dashed) lines. The image is not corrected by the primary beam. The vertical white bar indicates a 800~kpc size.}
\label{fig:noradio}
\end{figure}
%%%%%%%%%%%%%%%%%%%%%%%%%%%%%%%%%%%  End Figure 2
%%%%%%%%%%%%%%%%%%%%%%%%%%%%%%%%%%%  Begin Figure 2
\begin{figure}
\centering
\includegraphics[width=1\columnwidth]{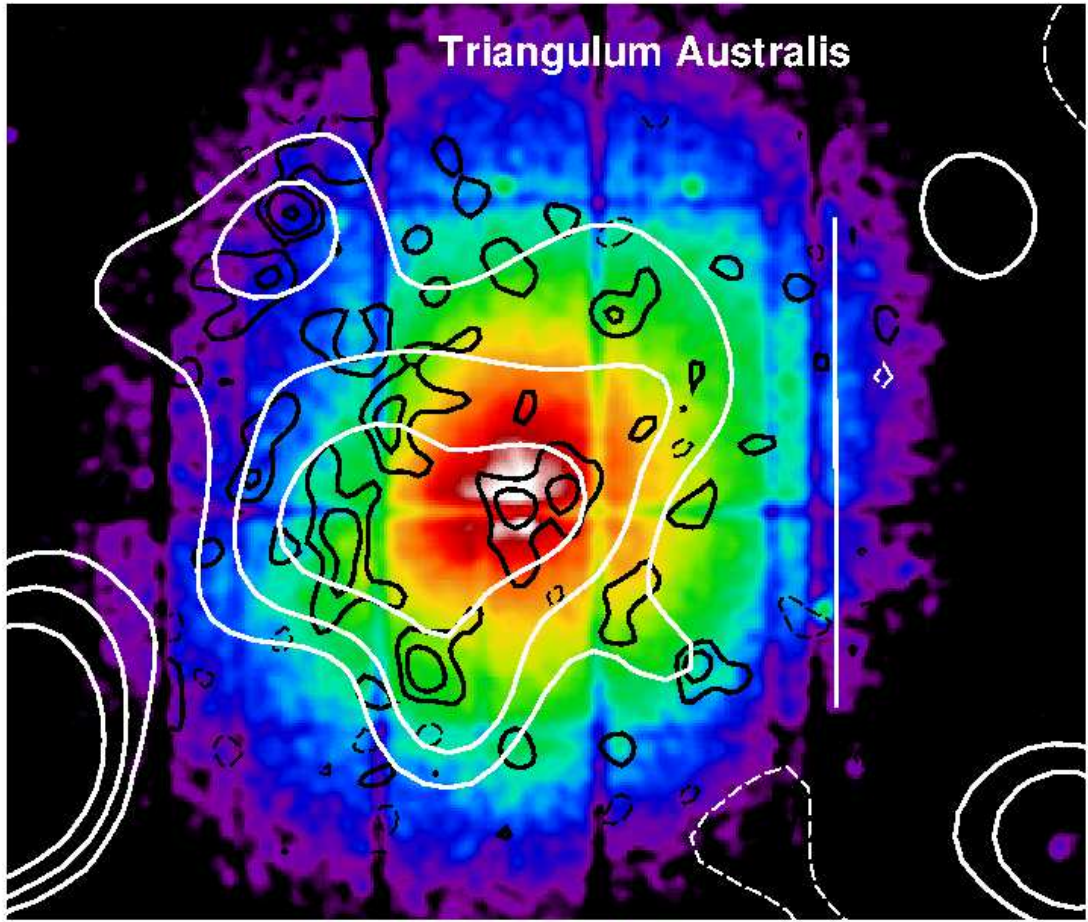}
\caption{1.86~GHz KAT--7 radio contours (white) of the Triangulum Australis on the {\it XMM--Newton} image (from Bernardi et al. 2016). Contours are drawn at $\pm$~2.5, 5, 10~mJy~beam$^{-1}$ and SUMSS contours (black) are drawn at $\pm$~2, 4, 8~mJy~beam$^{-1}$. The vertical white bar indicates a 800~kpc size. }
\label{fig:triangulum}
\end{figure}
%%%%%%%%%%%%%%%%%%%%%%%%%%%%%%%%%%%  End Figure 2
%%%%%%%%%%%%%%%%%%%%%%%%%%%%%%%%%%%  Begin Figure 2
\begin{figure}
\centering
\includegraphics[width=1\columnwidth]{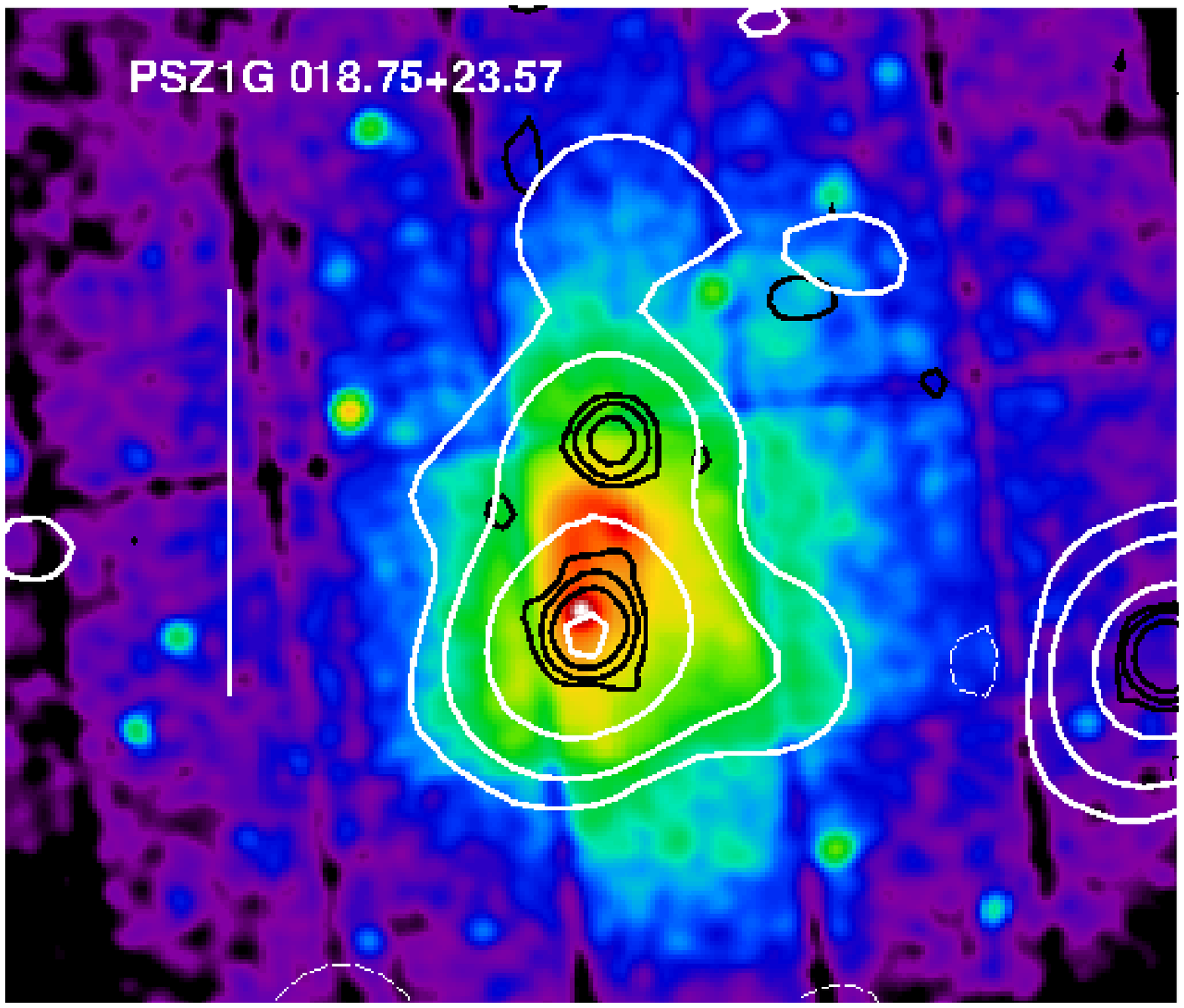}
\caption{1.86~GHz KAT--7 radio contours (white) of the PSZ1G\,018.75+23.57 cluster overlaid on the {\it XMM--Newton} image (from Bernardi et al. 2016). Contours are drawn at $\pm$~2.5, 10, 40, 160~mJy~beam$^{-1}$ and NVSS contours (black) are drawn at $\pm$~1.5, 6, 24~mJy~beam$^{-1}$. The vertical white bar indicates a 800~kpc size.}
\label{fig:planck}
\end{figure}
%%%%%%%%%%%%%%%%%%%%%%%%%%%%%%%%%%%  End Figure 2
%
Figure~\ref{fig:noradio} presents an example of a cluster where no diffuse emission was found down to the sensitivity level of the KAT--7 observations. Out of the entire sample, upper limits to the presence of a giant RH could be set for seven clusters which either did not show any emission in the KAT--7 data or for which radio emission could be attributed to compact sources and reliably subtracted using higher angular resolution literature data. The targets shown in Figure~\ref{fig:triangulum} and~\ref{fig:planck} are of particular interest as they represent the clearest RH candidate in our sample. In the case of the Triangulum Australis, we found emission extending over $\sim 15$~arcmin, corresponding to $\sim 900$~kpc, broadly consistent with what reported by Scaife et al. (2015), although the comparison with the SUMSS data indicates that there may be some contribution from compact sources. Figure~\ref{fig:planck} shows the case of PSZ1G\,018.75+23.57, where diffuse radio emission extends well beyond the contours of the two compact sources present in the NVSS image. Even after subtracting the NVSS sources, diffuse radio emission is still present over a 8-10~arcmin angular scale.\\
For six clusters of the sample we detected emission from compact radio sources and the KAT--7 limited angular resolution prevented a clear assessment of the presence of diffuse radio power potentially attributable to a RH. 

\section{Discussion}
\label{final_conclusions}

\begin{figure}
\centering
\includegraphics[width=0.9\columnwidth]{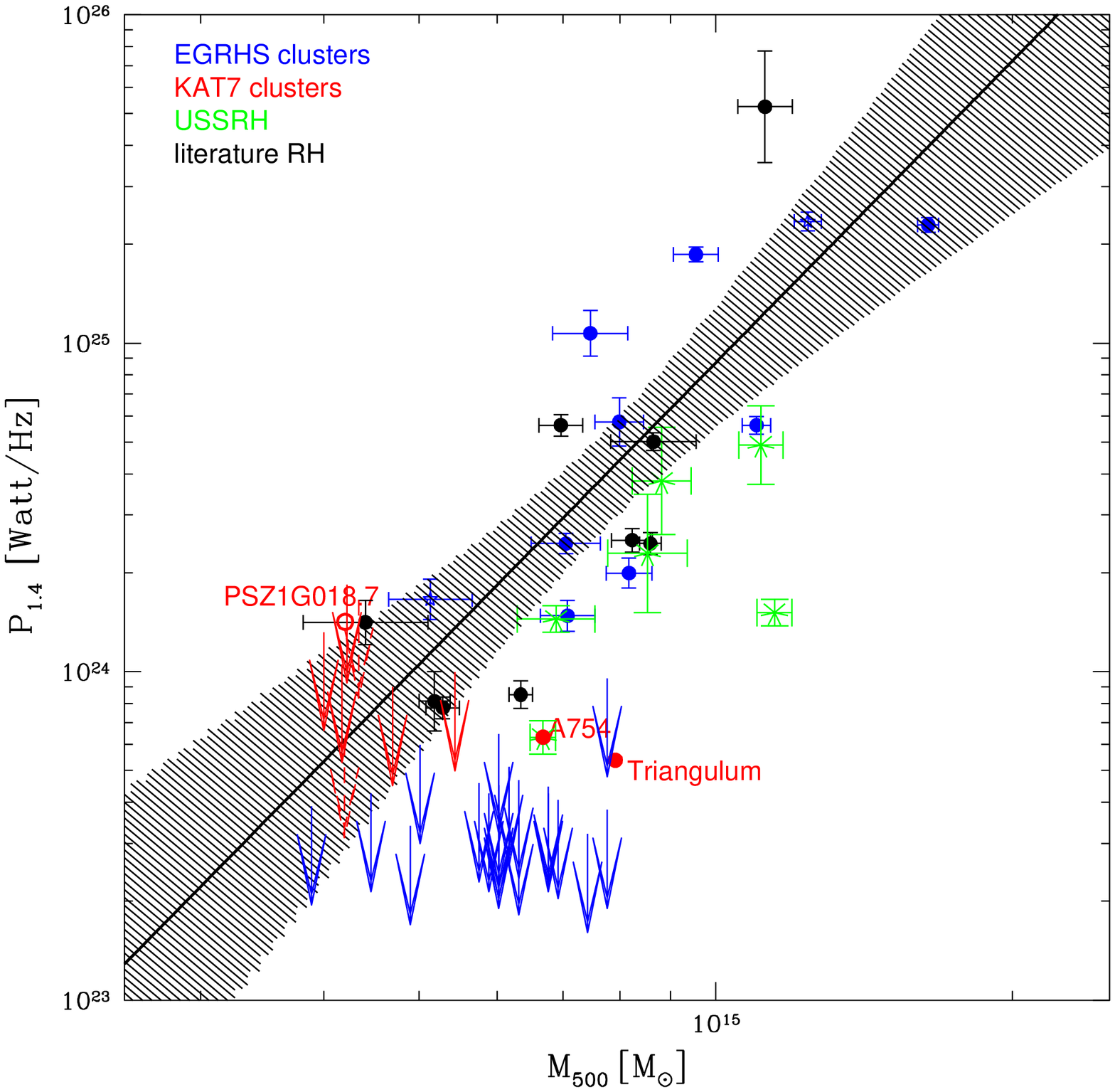}
\caption{Distribution of clusters in the P$_{\rm 1.4}-{\rm M}_{500}$ diagram (from Bernardi et al. 2016). 
Different symbols indicate RHs belonging to the Extended GMRT radio halo survey (EGRHS, blue dots, Venturi et al. 2007, 2008, Kale et al. 2013, 2015), RHs from the literature (black dots), ultra steep spectrum RHs (USSRH, green asterisks); upper limits from the EGRHS (blue arrows) and upper limits from KAT--7 observations (red arrows). Red, filled circles report the measured RH flux densities for the Triangulum Australis (re-analysis of the Scaife et al., 2015, data) and A\,754 (Macario et al. 2011) respectively. A\,754 was observed with KAT--7 as its RH is already know from theliterature. The PSZ1G\,018.75+23.57 flux density of the candidate RH (this work) after point sources were subtracted is reported as red, open circle. The best-fit relation to giant RHs only (black line) and its 95\% confidence region (shadowed region) are shown.}
\label{fig:LrM500}
\end{figure}

Upper limits to the flux density of a possible giant RH were derived using the injection methods (Venturi et al. 2008, Kale et al. 2013) for the seven targets clearly void of emission. They are reported in Figure~\ref{fig:LrM500} that includes the most updated compilation of RH measurements and upper limits and the best fit to the P$_{\rm 1.4}-{\rm M}_{500}$ correlation (from Bernardi et al. 2016). 

The KAT--7 data offer the first statistical information about cluster RHs in the M~$ > 4 \times 10^{14}$~M$_{\odot}$ range and confirm the lack of bright RHs in such systems, indicating that RHs more powerful than expected from the correlation must be rare which is in line with the predictions of the turbulent re-acceleration model (Cassano \& Brunetti 2005). Under the assumption that the P$_{\rm 1.4}-{\rm M}_{500}$ holds for smaller systems too, our results indicate that this correlation has a steep slope, of the form P$_{1.4} \propto {\rm M}_{500}^{\beta}$, with $\beta \ge 3$. Further observations at higher angular resolution and lower sensitivity will be key to probe the low end of the P$_{\rm 1.4}-{\rm M}_{500}$ correlation and provide more stringent tests of the RH formation models.
%
%%%%%%%%%%%%%%%%%%%%%%%% Begin Figure 7
%%%%%%%%%%%%%%%%%%%%%%  End of Figure 7

\section*{Acknowledgments}
This work is based on research supported by the National Research Foundation under grant 92725. Any opinion, finding and conclusion or recommendation expressed in this material is that of the author(s) and the NRF does not accept any liability in this regard. This work was also partly supported by the Executive Programme of Scientific and Technological Co-operation between the Italian Republic and the Republic of South Africa 2014--2016. The KAT--7 is supported by SKA South Africa and by the National Science Foundation of South Africa. Participation to the Conference was supported by the Ministry of Foreign Affairs and International Cooperation, Directorate General for the Country Promotion (Bilateral Grant Agreement ZA14GR02 - Mapping the Universe on the Pathway to SKA).

\end{document}